\documentclass{procs9x6}

\begin{document}

\title{Measurement of muon induced neutron background at shallow
sites}

\author{J. Wolf}

\address{Institut f\"ur experimentelle Kernphysik\\
Universit\"at Karlsruhe\\
Postfach 3640, 76021 Karlsruhe, Germany\\
E-mail: Joachim.Wolf@ik.fzk.de}


\maketitle

\abstracts{Cosmic muon induced neutrons are a major source of
background for low count rate experiments like neutrino
oscillation or dark matter searches. Especially at shallow sites
these neutrons are the limiting factor for the ultimate
sensitivity of the measurement. Measurements of the neutron rate
and counter measures including active veto and passive shielding
of the detector are discussed for two neutrino oscillation
experiments at shallow sites: the KARMEN accelerator based
experiment at RAL and the PALO VERDE reactor experiment.}

\section{Introduction}

\begin{figure}[ht]
\centerline{\epsfxsize=4.5in
            \epsfbox{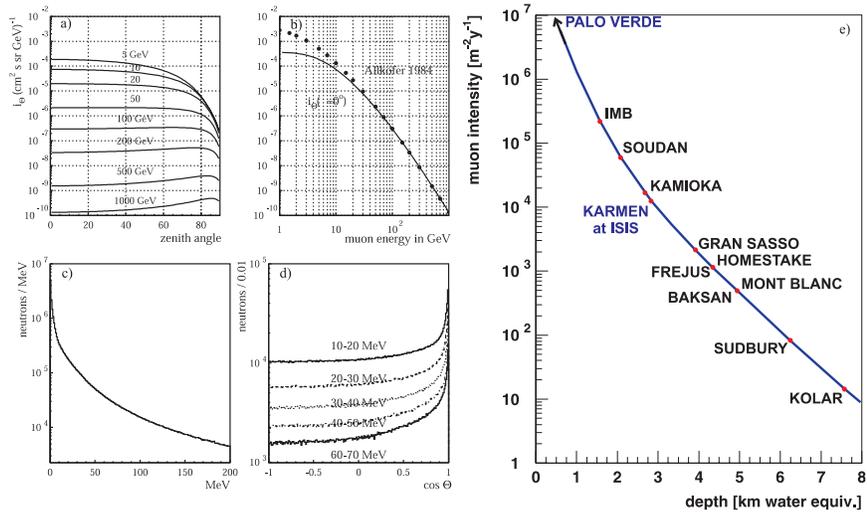}}
\caption{Cosmic ray muon flux and neutrons: a) muon flux at sea
level for different zenith angles and energy (simulation); b)
energy dependent muon flux at see level (dots are from
measurement$^1$); muon induced spallation neutrons (simulation):
c) energy spectrum, d) emission angle relative to incoming muon
track; e) muon flux as a function of depth (KARMEN: time structure
of ISIS equivalent to depth). \label{fig-01}}
\end{figure}

High energetic muons are, apart from neutrinos, the most
penetrating components of secondary cosmic radiation. They can be
detected even several thousand meters underground
(fig.~\ref{fig-01}e). In general, direct muon hits are suppressed
by active veto systems. However, high energy neutrons, produced by
muon interactions with nuclei outside the experimental setup, can
penetrate the veto system undetected. Experiments looking for rare
low energy events, like neutrino oscillation or dark matter
searches, can be dominated by this background. Two main reactions
of muons with nuclei can be distinguished\cite{ref_karm02}:
\begin{itemize}
\item {\bf$\mu^-$ capture:}
 Negative muons stopped in matter either decay ($\mu^- \rightarrow
 e^- + \nu_\mu + \bar\nu_e$) or are captured by a nucleus($\mu^- + p \rightarrow
 n + \nu_\mu$). The energy transferred to the nucleus in the process is
 between 15 and 20\,MeV and therefore above the neutron emission threshold.
\item {\bf Deep inelastic  scattering} (DIS) of muons on nuclei:
Virtual photons radiated from cosmic muons interact with a nucleus
and can produce one or more high energetic spallation neutrons
(see also\cite{ref_pv04}).
\end{itemize}
A detailed description of simulations of the muon flux and
subsequent muon induced neutron production have been described in
detail by Armbruster\cite{ref_karm02} and Wang et
al\cite{ref_pv04}. Figure~\ref{fig-01} shows the energy and zenith
angle dependence of the muon flux at sea level from a simulation
of the muon induced neutron background of the KARMEN neutrino
experiment. It also shows the simulated distribution of DIS
neutrons. The largest uncertainty of the simulation comes from the
total neutron yield of muon deep inelastic scattering.

The neutrino oscillation experiments KARMEN and Palo Verde both
searched for space- and time-correlated inverse beta decay
signatures from $\bar\nu_e$ events. The prompt positron from the
reaction $\bar\nu_e + p \rightarrow e^+ + n$ is followed by a
$\gamma$ cascade from neutron capture after a characteristic
thermalization time. Both detectors were based on organic liquid
scintillator for calorimetric measurement of the energy and a fast
time response. Gadolinium inside the detector was used to increase
the thermal capture cross section for neutrons.

Neutrons from deep inelastic scattering can penetrate into the
liquid scintillator, causing signals with visible energies up to
several hundred MeV through elastic n--p scattering (see
fig.~\ref{fig-01}c). After thermalization the neutrons are
captured, thus providing the sequential event signatures, which
are nearly identical to the inverse beta decay of a $\bar\nu_e$.
Since both detectors had no particle identification, they could
not distinguish between cosmic induced n-p recoil events and
positrons from $\bar\nu_e$ reactions.

\section{KARMEN}

The KARMEN\footnote{{\bf KA}rlsruhe {\bf R}utherford {\bf M}edium
{\bf E}nergy {\bf N}eutrino experiment} experiment was operated
from 1990 to 2001 at the ISIS spallation neutron source at the
Rutherford Appleton Laboratory in England. ISIS produces equal
fluxes of $\rm \nu_\mu$, $\rm \bar\nu_\mu$ and $\rm \nu_e$ from
$\pi^+$ decay and subsequent $\mu^+$ decay at rest. The unique
time structure of the ISIS neutrino pulses (10\,$\mu$s, 20\,Hz
repetition rate) provided a suppression of the continuous cosmic
muon flux for KARMEN, equivalent to a shielding of almost
3000\,{\em mwe} (meter water equivalent). KARMEN investigated
neutrino-nucleus interactions and searched for $\rm \bar\nu_\mu -
\bar\nu_e$ appearance oscillations\cite{ref_karm01}.

\begin{figure}[ht]
\centerline{\epsfxsize=4.5in
            \epsfbox{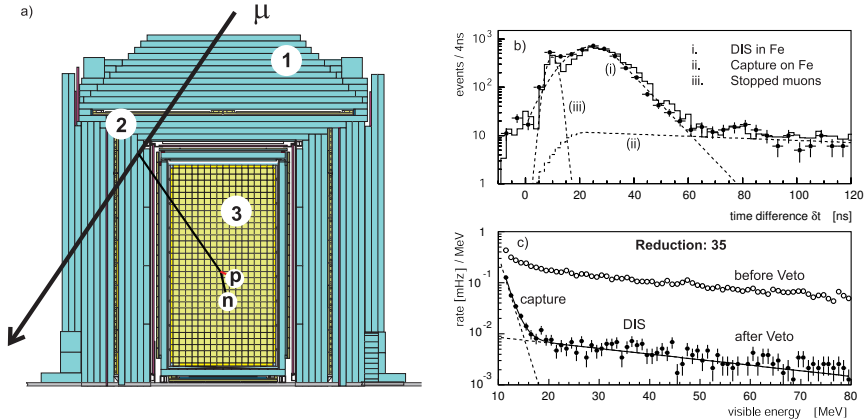}}
\caption{a) KARMEN detector (3) in a 7000 ton steel bunker (1) and
active veto system (2); muon induced neutrons: b) time of flight
spectrum, c) {\em visible} energy spectrum. \label{fig-03}}
\end{figure}

The KARMEN detector\cite{Dre90} was a segmented high resolution
liquid scintillation calorimeter (see Fig. \ref{fig-03}a). Its
volume (65\,m$^3$) was optically separated into 512 independent
modules. The walls of each cell contained Gadolinium coated paper
for an efficient detection of thermal neutrons. Two layers of veto
counters surrounded the scintillator tank. A blockhouse of 7000\,t
of steel slabs provided a passive shielding against the hadronic
and electro-magnetic component of cosmic showers. The modular
structure of the blockhouse allowed the integration of an
additional outer veto system half way through the steel shielding.
This veto system was installed in 1996, marking the beginning of
the KARMEN-2 experiment. The upgrade of the experimental
configuration reduced the cosmic induced background for the $\rm
\bar\nu_\mu - \bar\nu_e$
search\cite{ref_karm01}$^,$\cite{ref_karm02} to a level where beam
correlated neutrino-nucleus reactions dominated the background
rate.

Figure~\ref{fig-03}b shows the time correlation of muons tagged by
the outer veto system and the n-p recoil event in the central
detector. A subsequent neutron capture signal identified the event
as a muon induced neutron. The time distribution shows three
distinct contributions: i) DIS neutrons, ii) muon capture on iron
and iii) stopped muons in the detector. The solid histogram
represents the expected time distribution from a {\small
GEANT3.21} simulation, which is in good agreement with the
experimental data. Figure~\ref{fig-03}c shows the {\em visible}
energy distribution in the central detector of prompt events with
a correlated neutron capture signal (open circles). After removing
all events with an outer veto tag (full circles), the background
rate could be reduced by a factor of 35. The remaining spectrum
consists of two components. The soft component is caused by
neutrons from muon capture reactions and can be described as an
exponential distribution $e^{-E/E_0}$ with $E_0\approx 1.4$~MeV.
The much harder component attributed to DIS neutrons has an
exponential parameter of $E_0\approx 42$~MeV.

\section{Palo Verde}

\begin{figure}[ht]
\centerline{\epsfxsize=4.5in
            \epsfbox{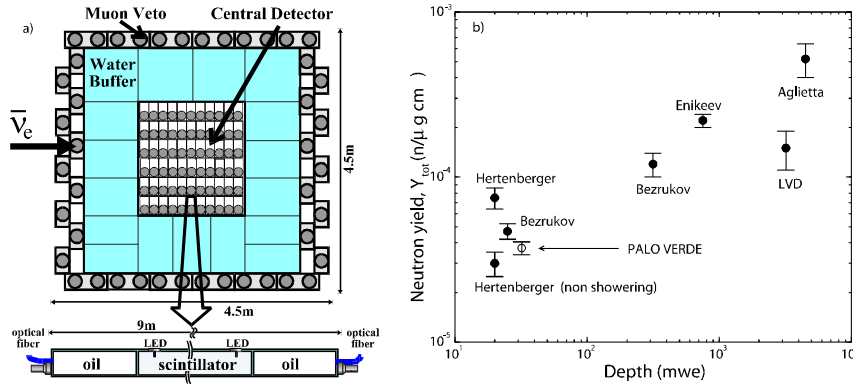}}
\caption{a) Palo Verde detector; b) neutron yield per muon as a
function of depth. \label{fig-04}}
\end{figure}

The Palo Verde neutrino oscillation experiment\cite{ref_pv01} was
operated from 1998 to 2000 at the Palo Verde Nuclear Generating
Station (11.6~GW) near Phoenix, Arizona. The main goal was the
search for $\rm \bar\nu_e - \bar\nu_x$ oscillations in the
disappearance mode. The detector was located in a shallow
underground site (32~{\em mwe} overburden), thus eliminating the
hadronic and electro-magnetic component of cosmic radiation and
reducing the muon flux by a factor of $\sim 5$. The segmented
detector contained 11.3\,tons of 0.1\% Gd-loaded liquid
scintillator in an array of 66 acrylic cells, as shown in
Fig.~\ref{fig-04}a. The central detector was surrounded by a 1\,m
water shield to moderate and capture background neutrons produced
by muons outside the detector and to absorb $\gamma$'s from the
laboratory walls. Surrounding the water tanks was a $4\pi$ veto
system made of 40 large liquid scintillator counters.

The $\rm \bar\nu_e$ events were identified by space- and
time-correlated $\rm e^+$ and n signals from an inverse
$\beta$-decay reaction. The visible energy of each sub-event was
in the range of 1 to 8 MeV\cite{ref_pv02}. The total sequential
event rate after veto and after all cuts was $\approx
50$\,events/day, including $\approx 25$ neutrino events per day.
More than half of the remaining background events came from
neutrons produced by deep inelastic muon scattering (DIS) outside
the veto system in the surrounding walls of the laboratory. The
sequence was produced either by n-p recoil in the scintillator
followed by neutron capture or by multi-neutron capture events.
Other background contributions came from unvetoed muons in the
water buffer and from random coincidences (mainly radioactivity).
The neutron background was identified and removed by a method
described in \cite{ref_pv05}.

A special measurement with a modified neutrino trigger was
performed\cite{ref_pv03} to determine the total neutron yield for
muons scattering inside the scintillator of the central detector.
Muon capture events were excluded by requiring through-going muon
tracks with at least two veto hits. Detection efficiencies, event
topologies and corrections for neutrons produced outside the
central detector were simulated, using {\small GEANT3.21}.
Hadronic interactions were simulated with FLUKA, low energetic
neutron transport with GCALOR. The analysis took into account the
production of up to three neutrons. Higher neutron multiplicities
were neglected. The resulting total neutron yield
\begin{equation*}
 Y_{tot} = \sum_{n=1}^{3} n\cdot Y_n =
 (3.60 \pm 0.09_{stat} \pm 0.31_{syst})\cdot 10^{-5}
 \frac{\rm neutrons}{\rm \mu \cdot g \cdot cm^{-2}}
\end{equation*}
is consistent with existing results shown in figure~\ref{fig-04}b.

\section{Conclusion}

Muon induced neutrons were the dominant part of cosmic ray
background for the neutrino oscillation searches with KARMEN and
Palo Verde. A good understanding of neutron production and
propagation in the vicinity of the detector has been demonstrated
by both experiments, showing good agreement between simulated and
measured neutron spectra in an MeV energy range. Apart from direct
particle identification (e.g. pulse shape discrimination), which
could not be used in this case, three counter-measures have been
taken to reduce the cosmic muon induced neutron background:
\begin{enumerate}
\item {\bf External passive shielding} of the muon flux by adding massive
shielding or going underground.
\item {\bf An active veto system} with high detection efficiency for muon
tracks in the vicinity of the detector.
\item {\bf Internal passive shielding} between veto system and
detector to remove neutrons produced by muons passing outside the
active veto system.
\end{enumerate}
As demonstrated by the successful upgrade of the KARMEN-2
experiment, a sufficient reduction of the neutron background can
only be achieved with a combination of an active veto system and
internal passive shielding.

\end{document}